\documentclass[aps,pre,twocolumn,superscriptaddress,showpacs,showkeys ]{revtex4-1}

\usepackage{graphicx}                   
\usepackage{hyperref}                   
\usepackage{amsmath,amssymb}            
\usepackage{epstopdf}
\usepackage[lofdepth,lotdepth]{subfig}

\usepackage{color}
\definecolor{orange}{rgb}{1,0.5,0}
\definecolor{brown}{rgb}{0.65, 0.16, 0.16}
\definecolor{phlox}{rgb}{0.87, 0.0, 1.0}

\begin{document}

\title{Invasion Sandpile Model}

\author{M. N. Najafi}
\affiliation{Department of Physics, University of Mohaghegh Ardabili, P.O. Box 179, Ardabil, Iran}
\email{morteza.nattagh@gmail.com}

\author{Z. Moghaddam}
\affiliation{Department of Physics, University of Mohaghegh Ardabili, P.O. Box 179, Ardabil, Iran}

\author{M. Samadpour}
\affiliation{Physics Department, K. N. Toosi University of Technology, Tehran, Iran}

\author{Nuno A. M. Ara\'ujo}
\affiliation{Departamento de F\'{\i}sica, Faculdade de Ci\^{e}ncias, Universidade de Lisboa, 
	1749-016 Lisboa, Portugal}
\affiliation{Centro de F\'{i}sica Te\'{o}rica e Computacional, Universidade de Lisboa, 
	1749-016 Lisboa, Portugal}

\begin{abstract}
Motivated by multiphase flow in reservoirs, we propose and study a two-species sandpile model in two dimensions. A pile of particles becomes unstable and topples if, at least one of the following two conditions is fulfilled: 1) the number of particles of one species in the pile exceeds a given threshold or 2) the total number of particles in the pile exceeds a second threshold. The latter mechanism leads to the invasion of one species through regions dominated by the other species. We studied numerically the statistics of the avalanches and identified two different regimes. For large avalanches the statistics is consistent with ordinary Bak-Tang-Weisenfeld model. Whereas, for small avalanches, we find a regime with different exponents. In particular, the fractal dimension of the external perimeter of avalanches is $D_f=1.47\pm 0.02$ and the exponent of their size distribution exponent is $\tau_s=0.95\pm 0.03$, which are significantly different from $D_f=1.25\pm 0.01$ and $\tau_s=1.26\pm 0.04$, observed for large avalanches.
\end{abstract}


\maketitle

\section{Introduction}
Invasion percolation (IP)~\cite{wilkinson1983invasion} is a standard model to study the dynamics of two immiscible phases (commonly denoted by wet and non-wet phases) in a porous medium~\cite{glass1996simulation,sheppard1999invasion}. During this process, the wet phase invades the non-wet phase, and the front separating the two fluids advances by invading the pore throat at the front with the lowest threshold~\cite{sheppard1999invasion}. This model provides valuable insight about the amount of invading fluid~\cite{wilkinson1983invasion}, the drying of capillary-porous material~\cite{prat1995isothermal}, and rock fracture networks~\cite{wettstein2012invasion}. However, in this simple model, features that are relevant to some practical applications are neglected. One example is the \textit{critical fluid saturation} (CFS) governing the dynamics of the fluid in the oil reservoirs~\cite{najafi2016water,najafi2014bak1}. In real porous media the fluid in a small region (comprised of many pores) is static and does not (macroscopically) move to the neighboring regions, until the accumulated water saturation in that region exceeds a certain saturation $(S_c)$, known as CFS~\cite{blunt2001flow,najafi2016water}. Physics of this type of threshold phenomenon is usually well captured by the sandpile-like models~\cite{najafi2016water,najafi2014bak1,araujo2013getting}. In Ref.~\cite{najafi2016water} the fluid toppling was taken into account as the main building block, and the sandpile model~\cite{BTW1988Self} was considered on top of the critical percolation cluster which was designed to compare the results with the Darcy reservoir model~\cite{Chin2002Quant,najafi2015geometrical}. For details of the model see, e.g. Ref.~\cite{najafi2014bak1}. The model was shown to be consistent with the Darcy reservoir model (with the same set of critical exponents) on the critical percolation cluster. This model was developed for a single phase (i.e., one particle species). Here, we generalize it to study a more realistic two-phase flow and refer to this model as \textit{Invasion Sandpile Model}. By measuring several statistical observables, we find that while, for large avalanche sizes, the statistics of the avalanches is consistent with what was previously observed for the one-species (ordinary BTW) model, for small avalanches the statistics is different~\cite{Bak1987Self}.\\

The paper is organized as follows: In the next section we describe the model. Section~\ref{SEC:results} is devoted to numerical details and results and we draw some conclusions in Sec.~\ref{concl}.

\section{The model}\label{Definition}

\begin{figure}
	\centerline{\includegraphics[scale=.45]{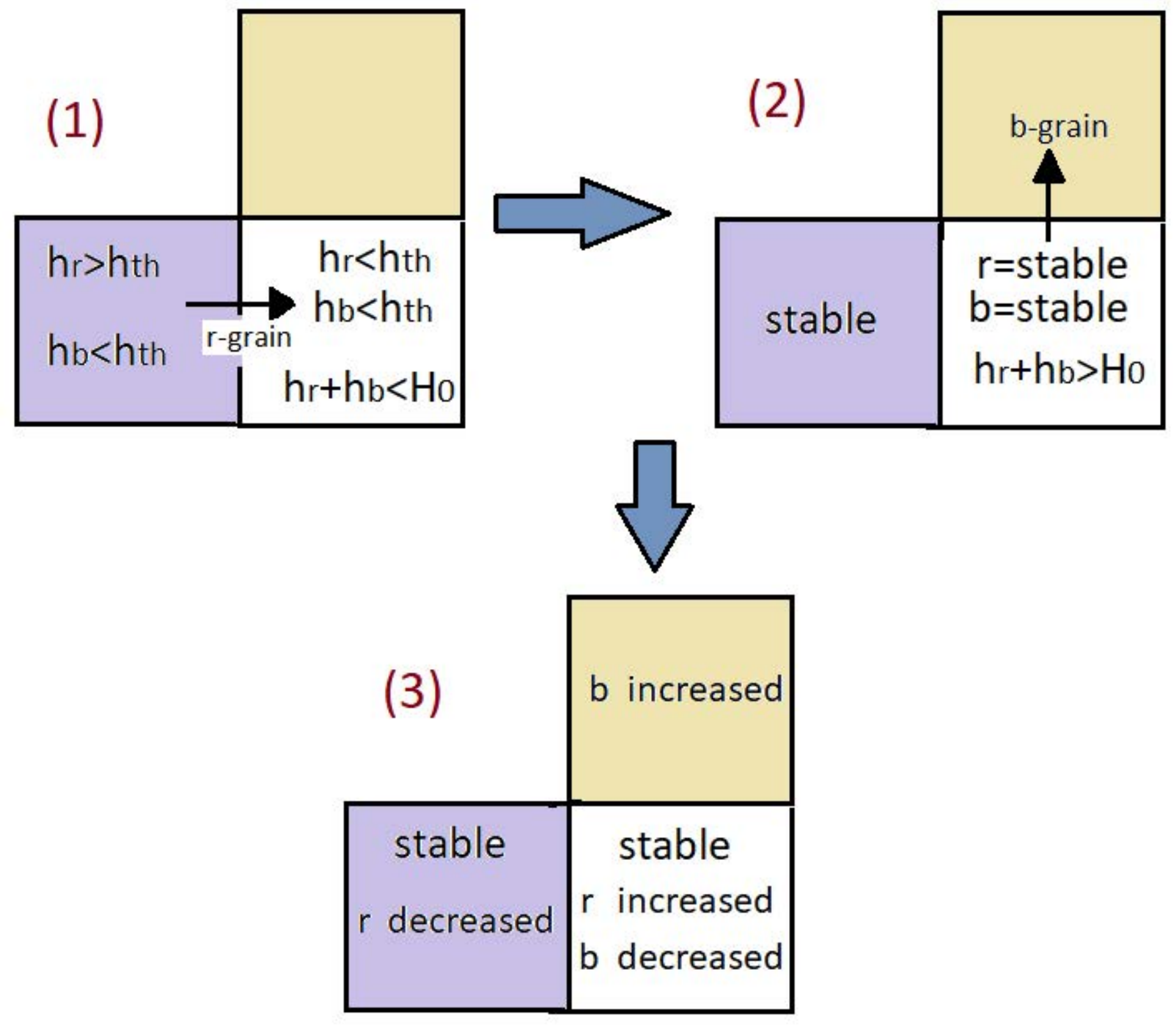}}
	\caption{Scheme with an example of invasion in the two-species sandpile model. In $(1)$ the left (blue) site becomes unstable since $h_r>h_{\text{th}}$. In $(2)$ however both grains are lower than $h_{\text{th}}$ for the right (white) site, but $h_r+h_b>H_0$. In this case, $r$ or $b$ is randomly chosen for toppling, here $b$ is chosen. Then in $(3)$ the $b$ content of the upper (dark yellow) site is increased. Therefore, effectively $r$ has invaded $b$ and pushed it towards another site.}
	\label{fig:Schematic}
\end{figure}

Let us consider a $L\times L$ square lattice and initially assign two random integers, $h_r$ and $h_b$, to each site, uniformly from the interval $\left\lbrace 1,2,3,..., h_{\text{th}}-1 \right\rbrace $, $h_{\text{th}}$ being the threshold for one species. $h_r$ and $h_b$ are the number of red and blue grains in our two-species sandpile model, representing the two (wet and non-wet) phases in the reservoir. The reported results are independent of the value of $ h_{\text{th}} $, so we set it to $20$. A site $i$ is considered stable if three conditions are fulfilled simultaneously. The first two are the standard ones for the one-species sandpile model, namely, $h_r(i)\leq h_\mathrm{th}$ and $h_b(i)\leq h_\mathrm{th}$, where $h_\mathrm{th}$ represents the CFS. The third one is $h_r(i)+h_b(i)\leq H_0$, where $H_0<2h_\mathrm{th}$ is the second threshold. This additional condition is motivated by the fact that in the non-linear Darcy equations, there is an auxiliary equation expressing that the sum of two phase saturations $S_w+S_o$ is a constant that depends on the capillary pressure. Thus, a site $i$ is \textit{unstable} and topples if at least one of the following conditions are met:\\
\\
C1: $h_r(i)>h_{\text{th}}$,\\
C2: $h_b(i)>h_{\text{th}}$,\\
C3: $h_r(i)+h_b(i)>H_0$\\
\\
 The dynamics goes as follows. Initially all $ h_r $ and $ h_b $ are chosen randomly from a uniform distribution, such that no site is unstable. Then, iteratively, we first choose a species (either $r$ or $b$, with equal probability) and a site $i$ at random to add a particle of that species, i.e. $h_x(i)\rightarrow h_x(i)+1$ where $x$ is the selected type. If that site becomes unstable, it topples, according to the following rule: If condition C1 is met, then $h_r(i)\rightarrow h_r(i)-1$ and $h_r(j)\rightarrow h_r(j)+1$ where $j$ is the neighbor of $i$ with the lowest red-grain content. If condition C2 is met, then $h_b(i)\rightarrow h_b(i)-1$ and $h_b(j)\rightarrow h_b(j)+1$ where $j$ is the neighbor of $i$ with the lowest blue-grain content. If condition C3 is met, then $h_x(i)\rightarrow h_x(i)-1$ and $h_x(j)\rightarrow h_x(j)+1$ where $j$ is the neighbor of $i$ with the lowest $x$-grain content, and $x$ is randomly chosen to be $r$ (red) or $b$ (blue). As a result of the relaxation of the original sites, the neighboring sites may become unstable and also topple. Therefore, the toppling process is repeated iteratively until all sites are stable again. This collective relaxation is called an \textit{avalanche}. The sand grains can leave the sample from the boundaries, just like in the ordinary BTW model~\cite{Bak1987Self}. Note that, with two species, an avalanche of one species might trigger an avalache of the other one, see example in Fig.~\ref{fig:Schematic} and details in the caption. The reason that we call this \textit{invasion} is that here one species pushes the other one due to the finite capacity of the pore, i.e. the total volume of the particles cannot exceed a threshold (see C3), as in real situations. In Darcy reservoir model, C3 is an auxiliary equation, where $H_0$ plays the role of the maximum finite saturation that is possible in a pore~\cite{najafi2016water,najafi2014bak1} and is the source of the invasion in the invasion percolation model~\cite{wilkinson1983invasion}.\\

In general, we find two types of avalanches: one-species and two-species avalanches. The first involve only the redistribution of grains of one species. In the second, there is mass transport of the two species. To analyze the dynamics, we measured the avalanche mass ($m$), defined as the total number of sites that toppled at least once and the avalanche size ($s$), which is the total number of topplings. We also define the avalanche cluster as the set of all sites that toppled at least once and analyzed the loop length ($l$) of its external perimeter, the mass gyration radius ($r_m\equiv\sqrt{\frac{1}{m}\sum_{i=1}^m\left|\textbf{r}_i-\bar{\textbf{r}}\right|^2 }$), and the loop gyration radius ($r\equiv\sqrt{\frac{1}{l}\sum_{i=1}^l\left|\textbf{r}_i-\bar{\textbf{r}}\right|^2 }$), where $\textbf{r}_i$ is the position of the $i$th site which has toppled at least once and $\bar{\textbf{r}}$ is the center of mass of the cluster. Note that the summation for the length runs over the sites in the external boundaries of the avalanche. We also measured the fractal dimension $ (D_f) $ of the loop of the external perimeter, using the relation $\left\langle \log l \right\rangle =D_f\left\langle \log r \right\rangle +$(constant). The other quantity that we investigate is the Green's function $G(i,j)$ for e.g. red grains, defined as follows: if the site $i$ is the site of injection of red sand grains, then $G(i,j)$ is the average number of topplings of red grains in site $j$. For the case that $i$ and $j$ are both distant from the boundaries, this function depends on the Euclidean distance between the sites~\cite{dhar1999some}. The same function is defined for the blue grains and also total avalanches. For the one-species BTW model, this function is logarithmic as expected from the free ghost field theory~\cite{najafi2012avalanche,Najafi2012Observation,najafi2018coupling}.

\section{Results}\label{SEC:results}
We considered square lattices of linear length $L=64,128,256,512,1024$ and $2048$. All statistical analyzes were performed in recurrent configurations. To reduce temporal correlations we average over $2\times 10^6$ samples of avalanches corresponding to every 100th avalanche in the time series. When we start from a random initial configuration (for both species), the average height (for total and each species) initially grows linearly but eventually saturates for long enough times. This is also the case for the ordinary BTW model, but the crossover time between regimes is much larger for the two species model. We analyze here the resulting avalanches in the long-time regime, in which the configurations are \textit{recurrent}.\\

Contrary to the ordinary BTW model, here the avalanches for each phase are non-local, i.e. the set of toppled red (or blue) sites are not necessarily connected due to invasion. Therefore, in general, the red (blue) avalanche is composed of some distinct smaller simply connected red (blue) avalanches, whereas the total avalanche is simply connected. Here we extracted the simply connected components using the Hoshen-Kopelman algorithm~\cite{hoshen1976percolation}.

\subsection{Two regimes and the crossover for one-species avalanches}
\begin{figure*}
	\centering
	\subfloat[][]{
		\includegraphics[width=0.45\textwidth]{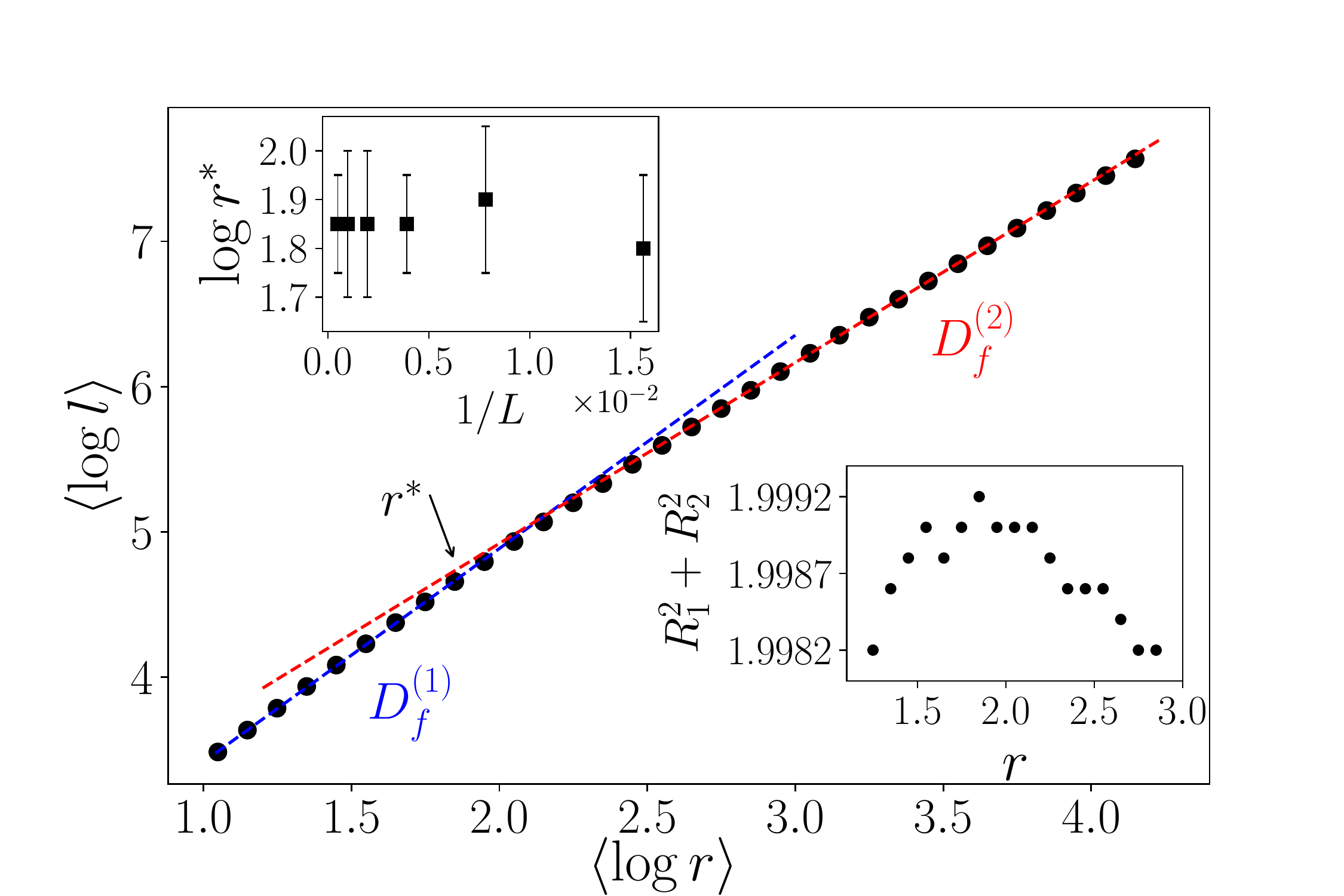}
		\label{fig:FD_red256}}
	\subfloat[][]{
		\includegraphics[width=0.45\textwidth]{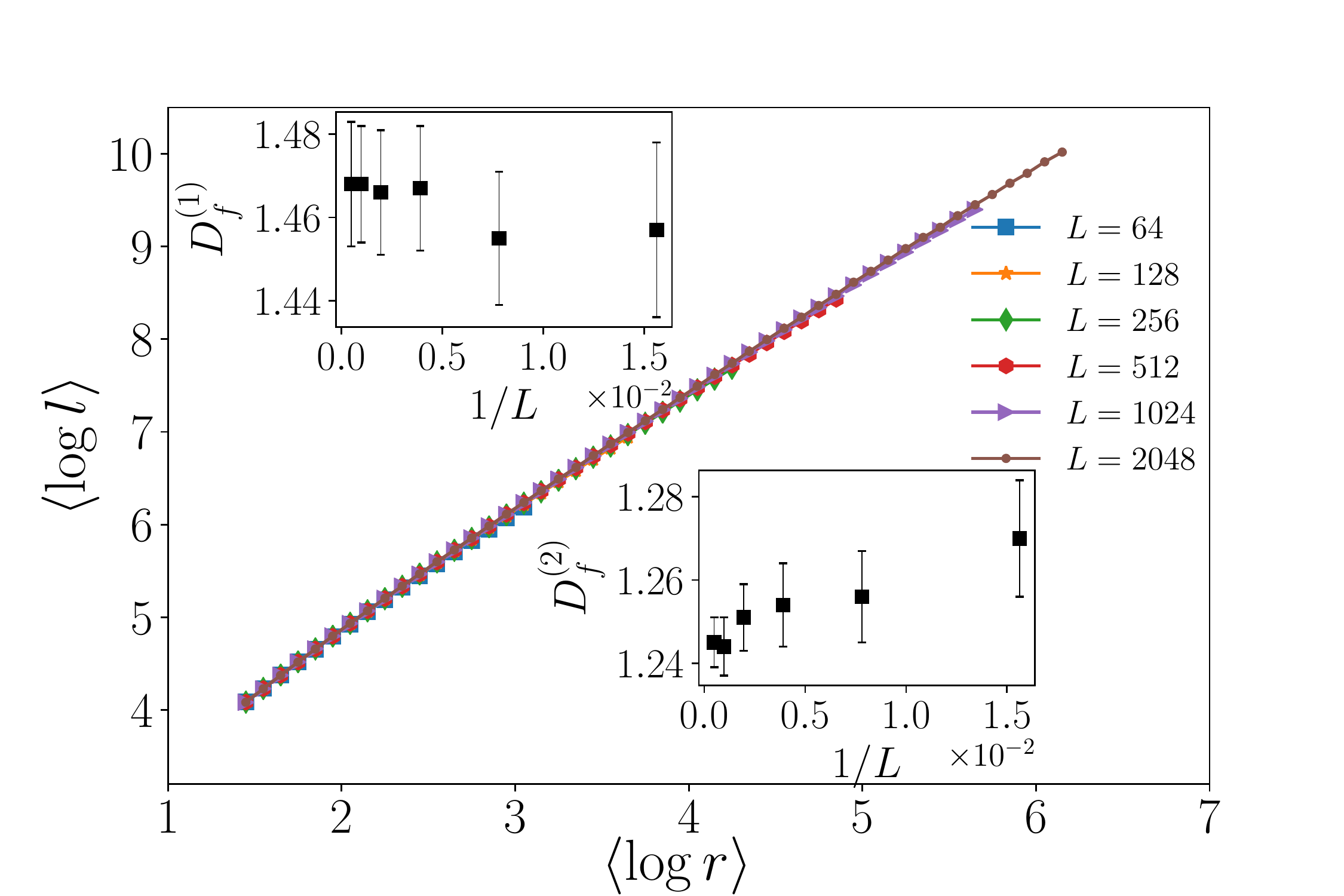}
		\label{fig:F_D_red}}
	\qquad
	\subfloat[][]{
		\includegraphics[width=0.45\textwidth]{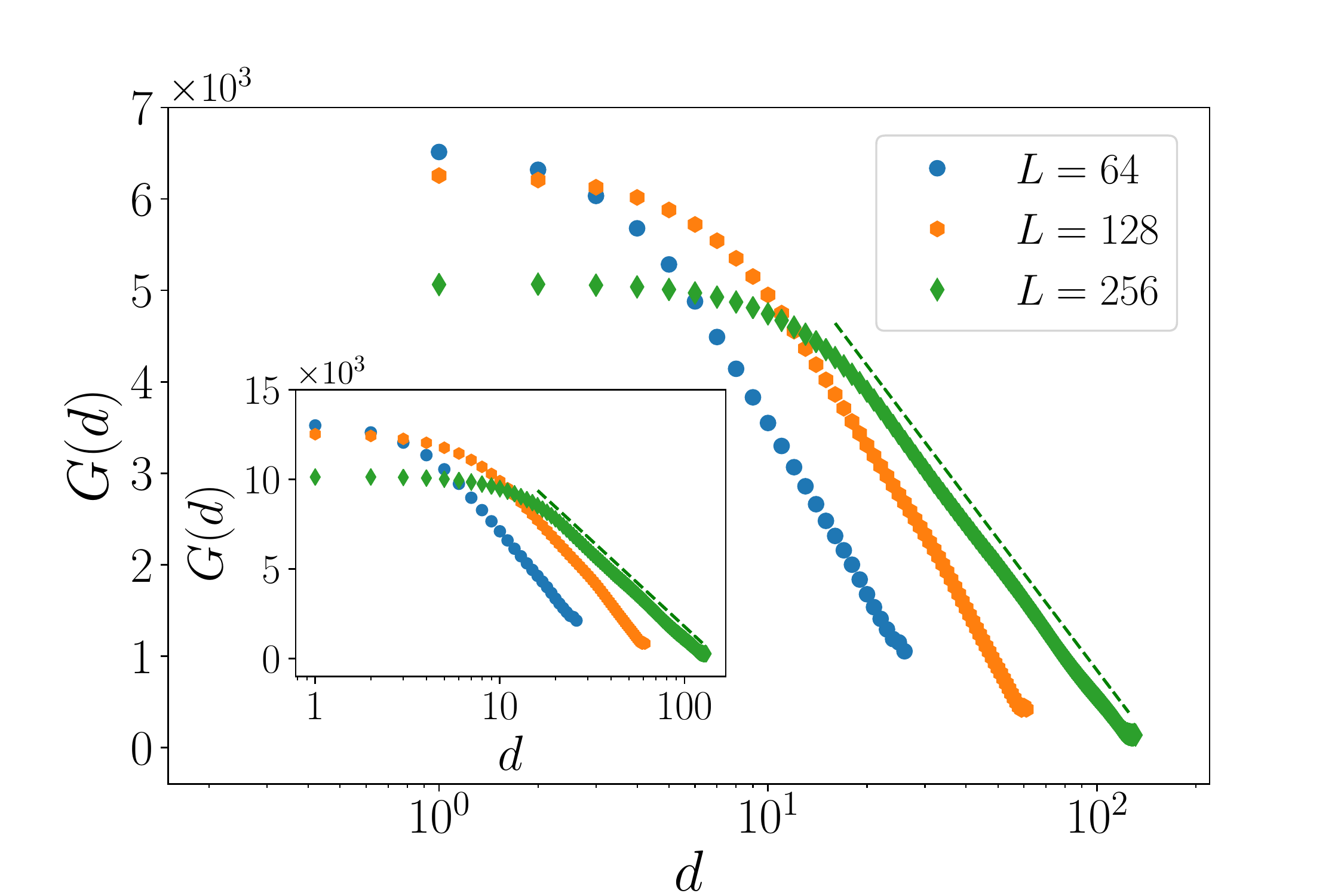}
		\label{fig:Green_red}}
	\subfloat[][]{
		\includegraphics[width=0.45\textwidth]{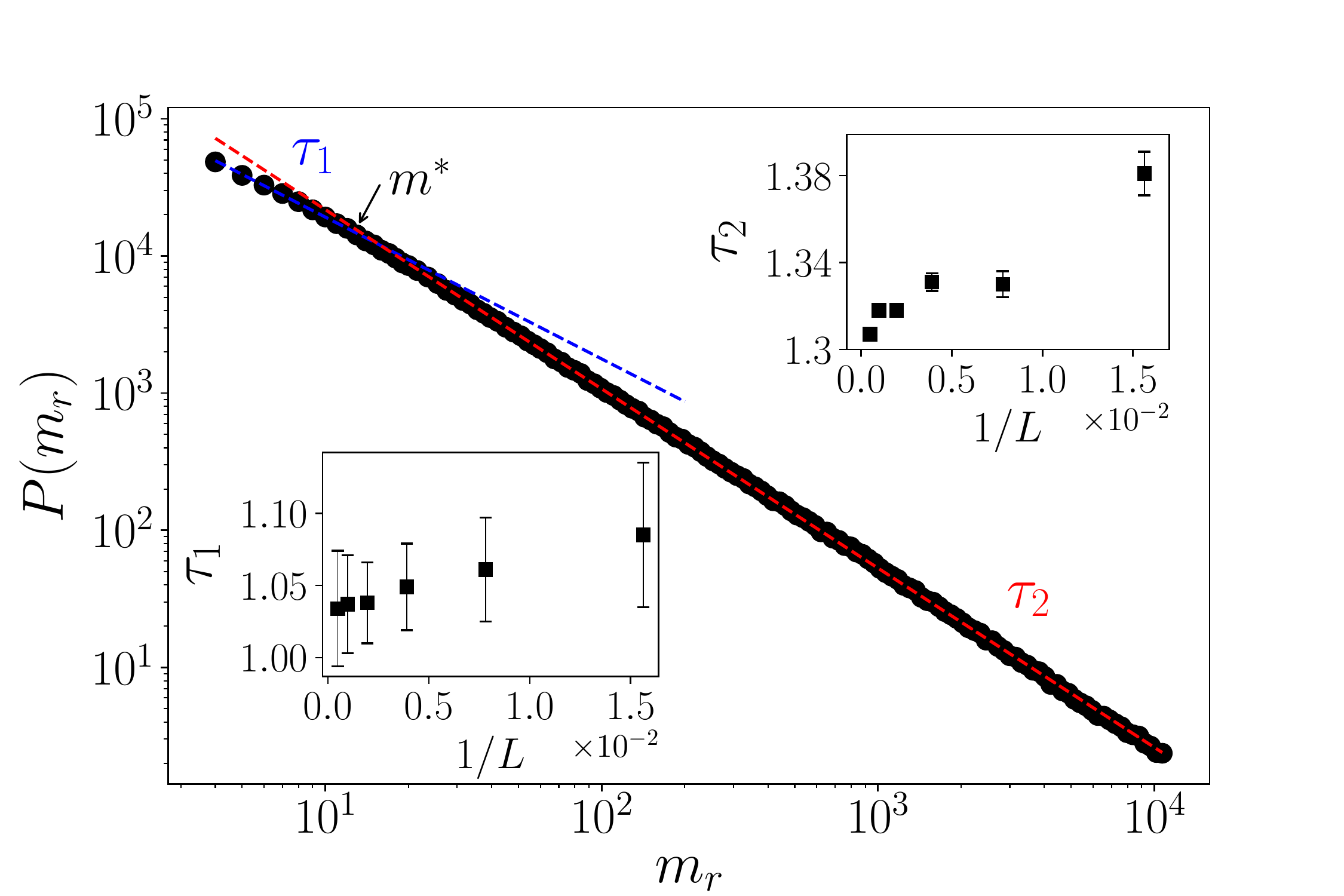}
		\label{fig:P(mass_red)}}
	\caption{(Color online): (a) The average $\left\langle \log l \right\rangle $ in terms of $\left\langle \log r \right\rangle $ for which the relation gives the fractal dimension $D_f$. $r^*$ is the separator of the small and large scale regimes. (b) $L$-dependent fractal dimension. $D_f^{(1)}$ is the fractal dimension for small scales (upper inset) and $D_f^{(2)}$ is the fractal dimension for large scales (lower inset). (c) The Green function in terms of distance for one species. Inset shows the same for the two-species avalanches. The function scales linearly in the semi-log plot for long enough distances (the straight lines are linear fits with $L$-dependent slopes). (d) The distribution function of avalanche mass ($m_r$) for one species ($r$ stands for red) with the corresponding crossover point. $\tau_2$ ($\tau_1$) is shown in the upper (lower) inset in terms of inverse system size $L^{-1}$.  }
	\label{Fig:first}
\end{figure*}

We discuss now the results for one-species avalanches, By symmetry, results for blue and red avalanches are equivalent. The fractal dimension $D_f$ is shown in Fig.~\ref{fig:FD_red256} and~\ref{fig:F_D_red} for which using a finite-size analysis (the insets of figure~\ref{fig:F_D_red} in which the exponents are plotted in terms of $1/L$, and the fractal dimensions are obtained by extrapolation) we clearly see two different regimes: for large avalanche sizes the fractal dimension $D_f^{(2)}$ is consistent with $\frac{5}{4}$ observed for the 2D BTW model~\cite{Lubeck1997BTW}. But for small avalanches, $D_f^{(1)}=1.47\pm 0.02$. Figure~\ref{fig:Green_red} shows also that for large avalanches, the Green's function is logarithmic, consistent with the BTW universality class~\cite{najafi2012avalanche}. These results suggest a crossover between two different regimes. \\

 To estimate the fractal dimensions in Fig.~\ref{fig:FD_red256} we extend the $R^2$ test~\cite{glantz1990primer} for two regimes to extract the crossover point (see the upper inset of Fig.~\ref{fig:FD_red256}), which is also relevant to obtain all the other exponents (see for example the insets of Fig.~\ref{fig:F_D_red}). For all measures, we find two distinct linear behaviors (in terms of $x$) with a crossover point in between ($x^*$, like $r^*$ in Fig.~\ref{fig:FD_red256}). For each value of $x^*$ we determine the $R^2$ of the fit of that regime (lower and higher than $x^*$), which obviously depend on $x^*$. Let us name the $R^2$ of the first and the second regimes as $R_1^2 (x^*)$ and $R_2^2 (x^*)$ respectively. For each regime, the higher the $R^2$, the better the fitting. Thus, in order to identify the crossover point, we find the fitting to both regimes that maximizes $R^2 (x^*)=R_1^2 (x^*)+R_2^2 (x^*)$. In the lower inset of Fig.~\ref{fig:FD_red256} we plot the $ R^2(r^*)$ for different values of $r^*$. We define the crossover point as the value of $r^*$ for which $R^2$ is a maximum, i.e. $r^*=7.1\pm 0.2$ for the $L=256$, and the error bar is obtained as usual for the least-squares method. \\

\begin{figure*}
	\centering
	\subfloat[][]{
		\includegraphics[width=0.31\textwidth]{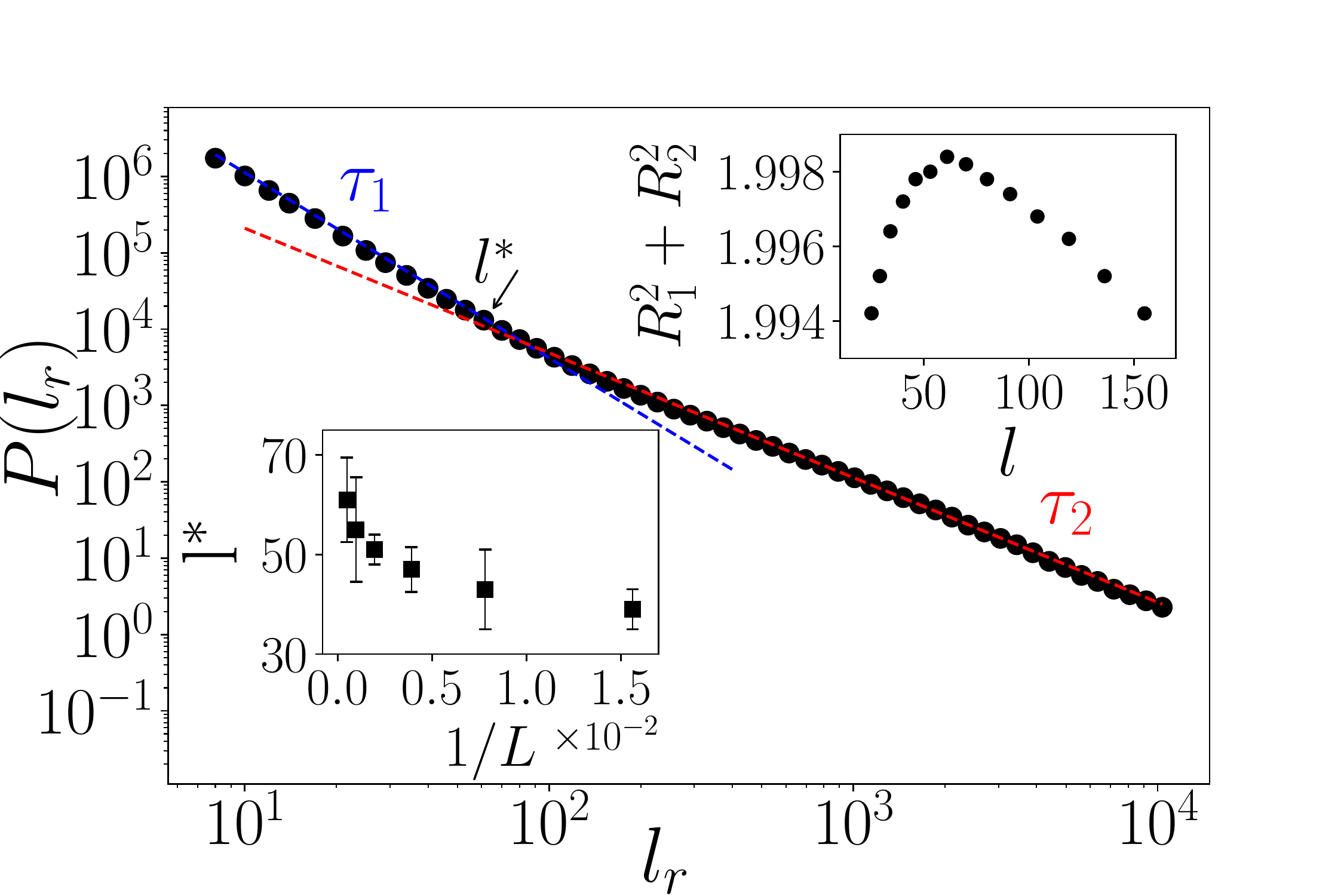}
		\label{fig:l_red2048}}
	\subfloat[][]{
		\includegraphics[width=0.31\textwidth]{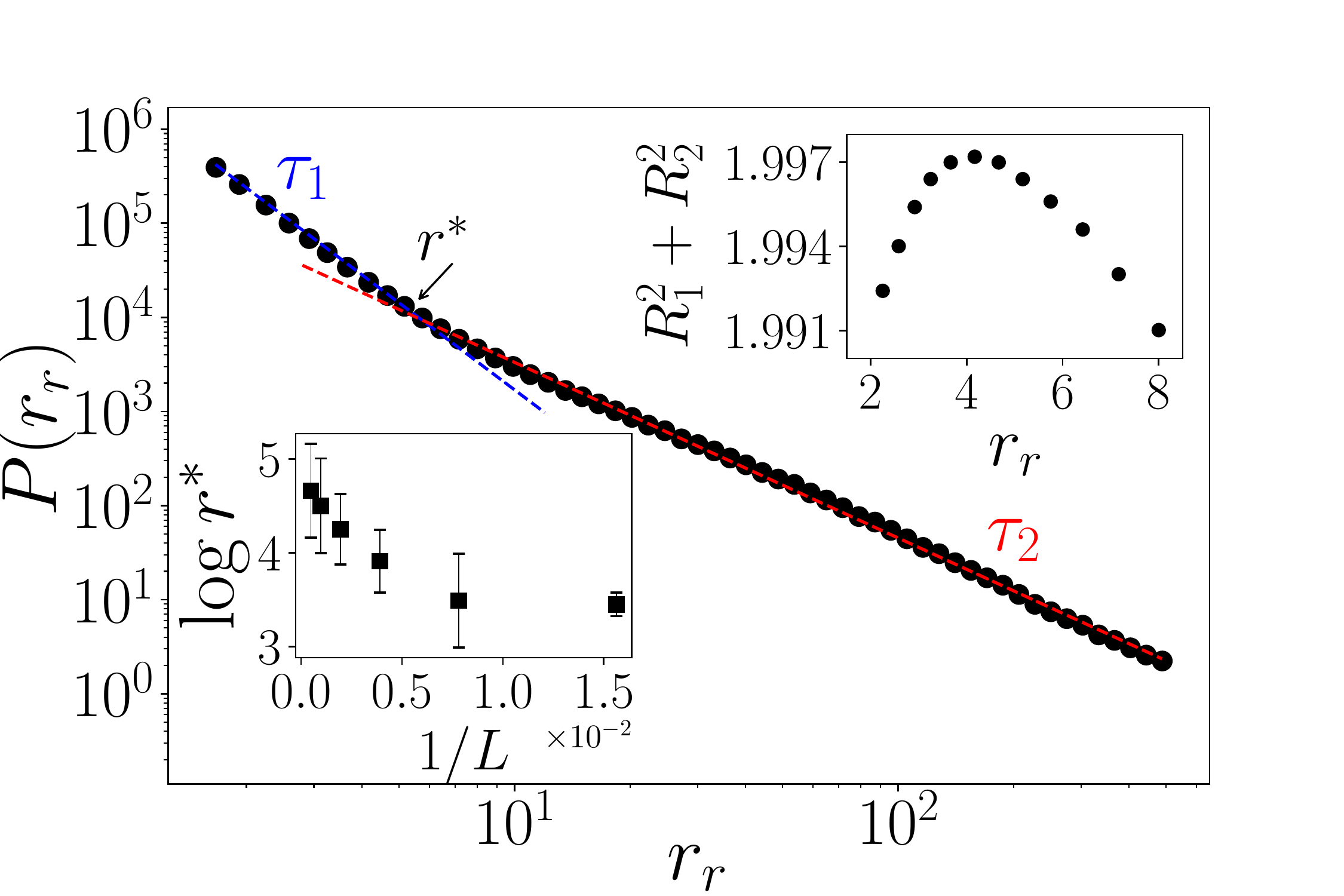}
		\label{fig:r_red2048}}
	\subfloat[][]{
		\includegraphics[width=0.31\textwidth]{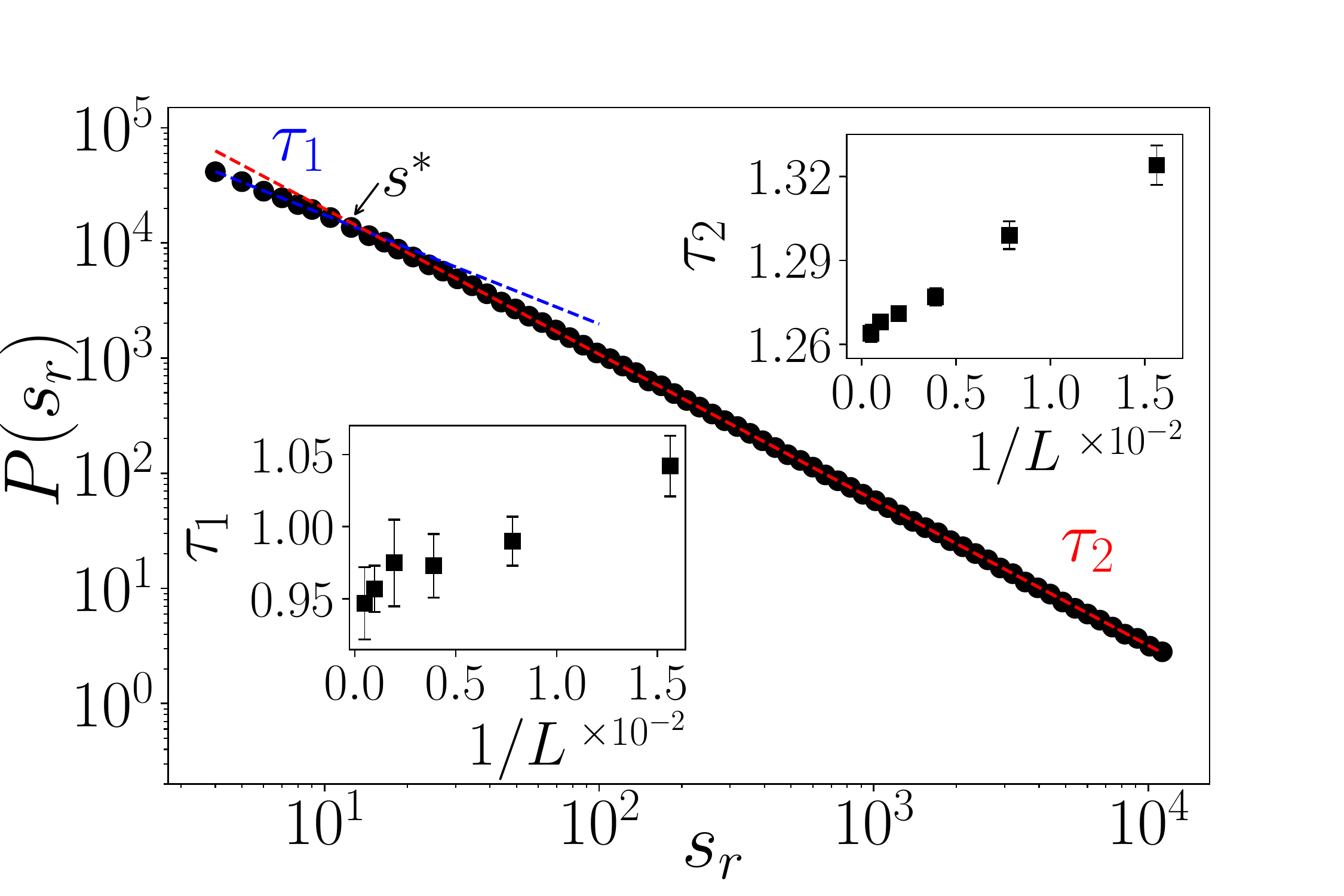}
		\label{fig:P(size_red)}}
	\qquad	
	\subfloat[][]{
		\includegraphics[width=0.45\textwidth]{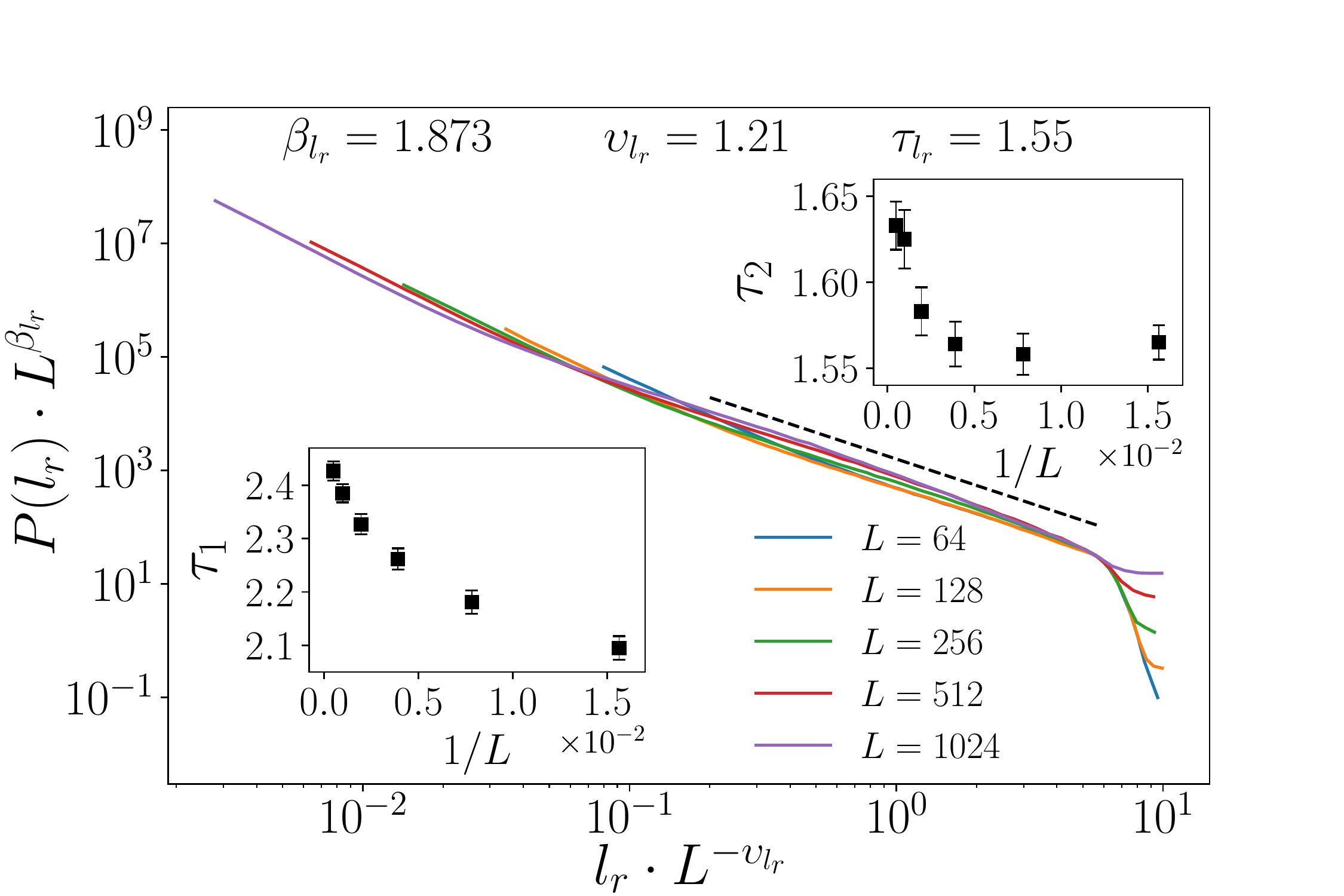}
		\label{fig:P(l_red)}}
	\subfloat[][]{
		\includegraphics[width=0.45\textwidth]{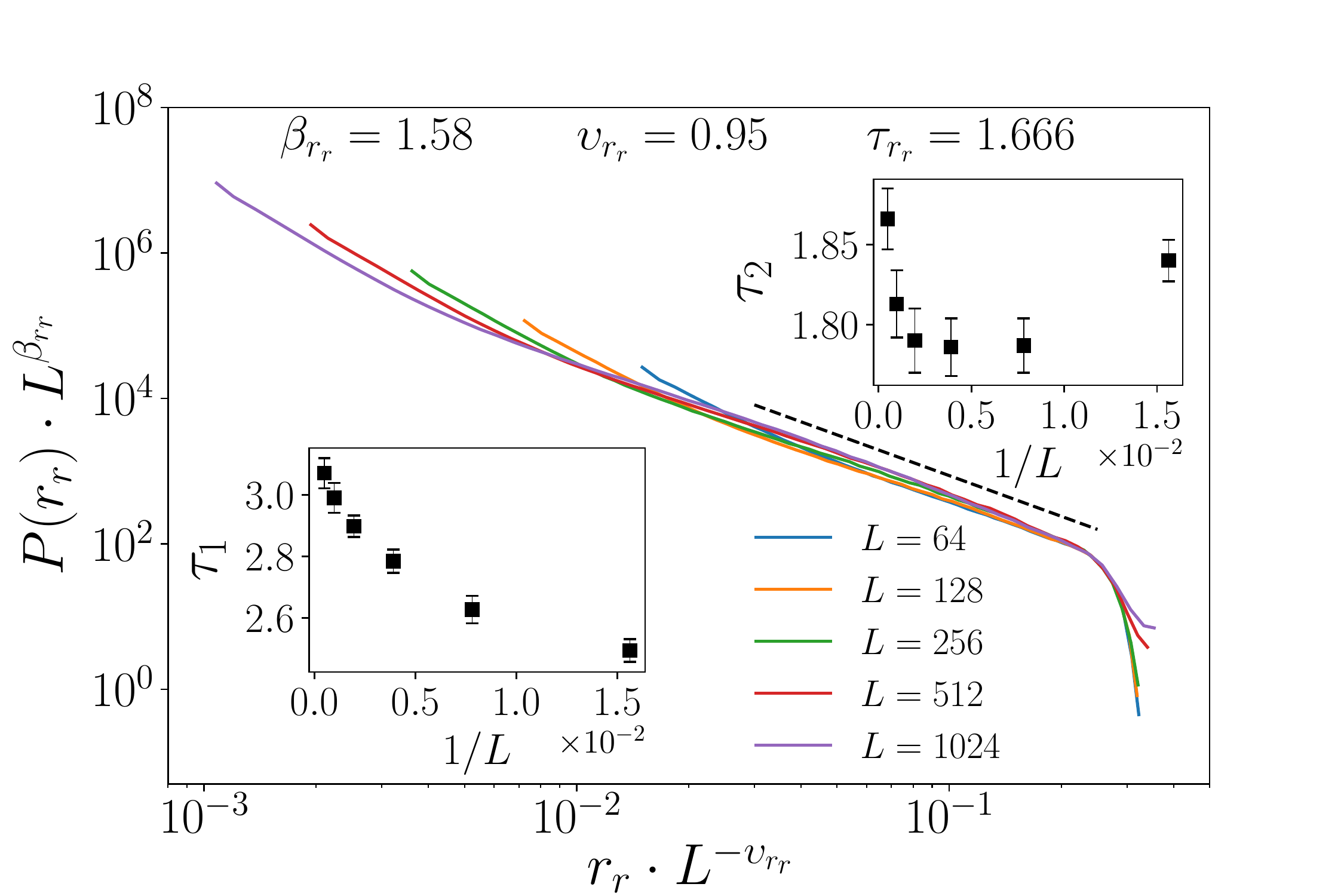}
		\label{fig:P(r_red)}}
	\caption{(Color online): The distribution function of (a) $l_r$ (red grains) and (b) $r$, along with the definition of the crossover points by plotting $R_1^2+R_2^2$ in terms of the tentative crossover point in the upper insets of the left panel. The crossover points are shown in terms of $L^{-1}$ in the lower insets of the left panel. In the right panel figures, the upper (lower) insets are $\tau_2$ ($\tau_1$) in terms of $L^{-1}$. (c) The distribution function of $s_r$, and the corresponding exponents $\tau_1$ and $\tau_2$ for small and large scales in the lower and upper insets respectively. The data collapse for the distribution function of (d) $l_r$ and (e) $r$.  }
	\label{Fig:results1}
\end{figure*}

Using the same strategy for all other quantities, e.g. for the avalanche mass distribution in Figure~\ref{fig:P(mass_red)}, we estimated a crossover point between the large and the small avalanche regimes. For large avalanches the exponent of the avalanche mass distribution is $\tau_2=1.32\pm 0.02$, which is in agreement with $\tau_m=4/3$ previously reported for the BTW model in 2D~\cite{Lubeck1997BTW,Lubeck1997Numerical,najafi2012avalanche}. However, for small avalanche sizes, the exponent $\tau_1=1.04\pm 0.04$ is different (see the insets). Figures~\ref{fig:l_red2048},~\ref{fig:r_red2048}, and~\ref{fig:P(size_red)} are the analyses for the distribution functions of loop length ($l$), gyration radius ($r$) and size ($s$). The amount of $r^*$ is compatible with the cross over point found for the fractal dimension. These figures reveal that the considered $x^*$s extrapolate to a finite value as $L\rightarrow\infty$, so that e.g. $\frac{r^*}{L}\rightarrow 0$. Therefore, we conclude that, in the thermodynamic limit the small-avalanche regime vanishes, i.e. the BTW-universality class is the only relevant one, in the thermodynamic limit. The exponents for small and large scales are shown in the insets of Figs.~\ref{fig:P(l_red)} and~\ref{fig:P(r_red)}, whereas in their main panels we show the data collapse for the large avalanche regime. This data collapse is based on the finite size scaling relation:
\begin{equation}
P(x)=x^{-\tau_x}F_x\left(\frac{x}{L^{\nu_x}} \right)=L^{-\beta_x}g_x\left(\frac{x}{L^{\nu_x}} \right) 
\end{equation}
where $x=l,s,m,r,r_m$, and $\tau_x$ and $\nu_x$ are its critical exponents, $\beta_x=\frac{\tau_x}{\nu_x}$ and $F(y)$ and $g_x(y)=y^{-\tau_x}F(y)$ is the universal function with the limits $\lim_{y\rightarrow 0}F(y)=$const., and $\lim_{y\rightarrow 0}g_x(y)\sim y^{-\tau}$.\\

All the obtained exponents are consistent with the BTW universality class, whereas the data for the small avalanches are completely different. The values of the exponents for the different regimes are summarized in TABLE~\ref{tab:exponents}, from which we observe compatible results for $\tau_2$ and $\tau^{\text{2D BTW}}$ over all calculated observables. We have observed that the avalanches with linear extension smaller than $r^*$ are often single component, whereas the number of the connected components are more than one for avalanches with larger extents. Therefore, we relate this crossover to the point where one goes from scales for which the avalanches are disconnected (non-local effects due to the interaction between the different species) to a regime where all avalanches are a single connected component. \\

\begin{table}
	\caption{The exponents $\beta$, $\nu$, $\tau_1$, and $\tau_2$ for $m$, $s$, $l$, and $r$ corresponding to the red avalanches. The last row contains the exponents for the 2D BTW model for the sake of comparison with $\tau_2(L\rightarrow \infty)$~\cite{Lubeck1997Numerical,najafi2012avalanche}.}
	\begin{tabular}{c | c | c | c | c }
		\hline quantity  & $m$ & $s$ & $l$  & $r$ \\
		\hline $\tau_1(L\rightarrow \infty)$  & $1.04\pm 0.04$ & $0.95\pm 0.03$ & $2.5\pm 0.03$  & $3.1\pm 0.1$ \\
		\hline $\tau_2(L\rightarrow \infty)$ & $1.32\pm 0.02$ & $1.26\pm 0.04$ & $1.63\pm 0.03$  & $1.8\pm 0.1$ \\
		\hline $\beta$   & $-$ & $-$ & $1.87\pm 0.05$  & $1.58\pm 0.03$ \\
		\hline $\nu$     & $-$ & $-$ & $1.21\pm 0.03$  & $0.95\pm 0.03$ \\
		\hline
		\hline $\tau^{\text{2D BTW}}$ & $1.33\pm 0.01$ & $1.29\pm 0.01$ & $1.25\pm 0.03$  & $1.66\pm 0.01$ \\
		\hline
	\end{tabular}
	\label{tab:exponents}
\end{table}

\subsection{The results for two-species avalanches}
\begin{figure}
	\centerline{\includegraphics[scale=.45]{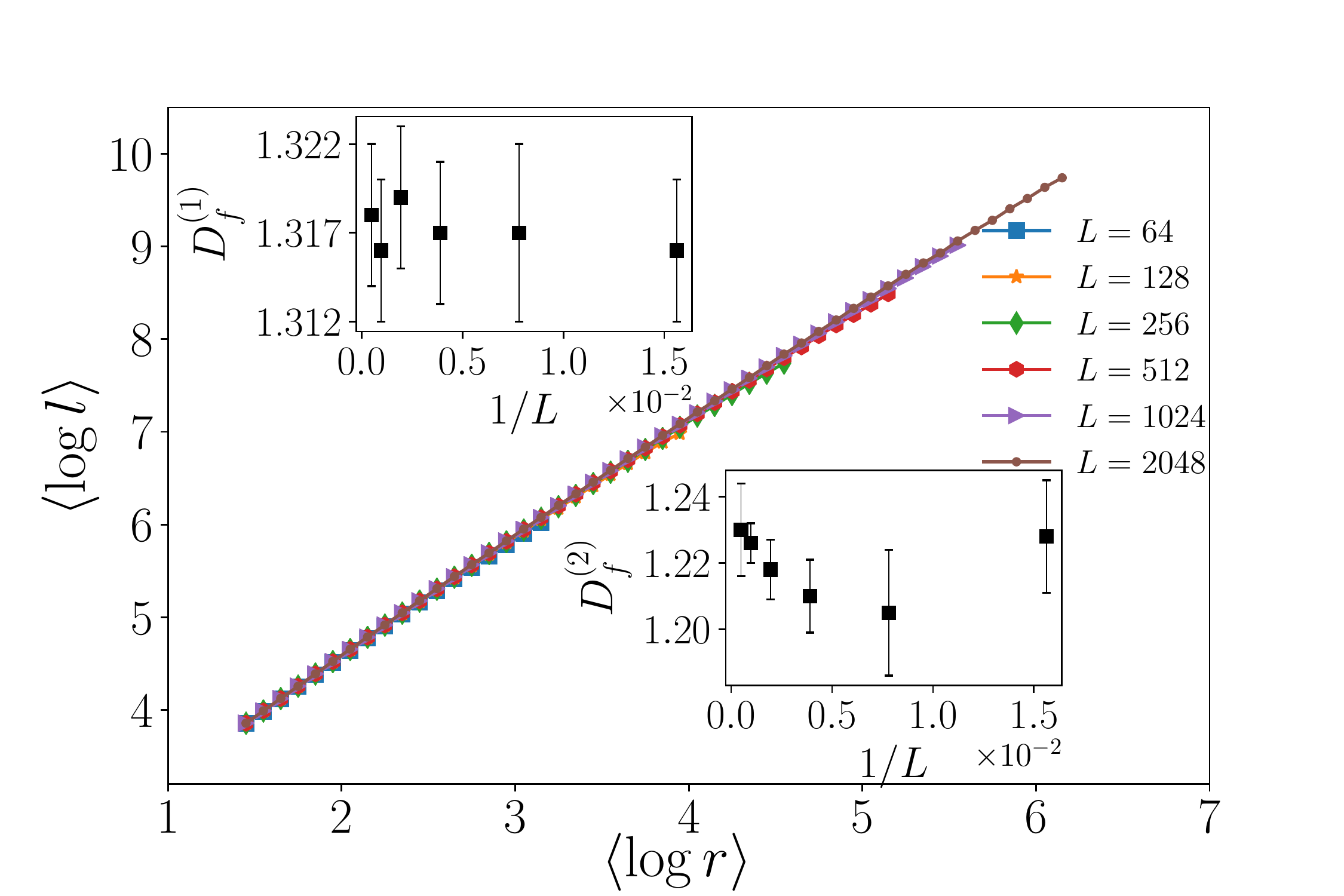}}
	\caption{The fractal dimension of two-species avalanches obtained from the scaling of $\left\langle\log l \right\rangle $-$\left\langle \log r\right\rangle $. Upper (lower) inset is the fractal dimension for small (large) avalanche regime, i.e. $D_f^{(1)}$ and $D_f^{(2)}$ respectively.}
	\label{fig:F_D_total}
\end{figure}

Let us now analyze the two-species avalanches. In the inset of Fig.~\ref{fig:Green_red} we plot the Green's function, which scales logarithmically with the distance, as in the case of one-species avalanches. Also from the Fig.~\ref{fig:F_D_total} for the fractal dimension we see that there are two regimes, and $D_f=1.24\pm 0.01$ for large avalanches consistent with 2D BTW model. However, the fractal dimension for small avalanches is $1.31\pm 0.01$, which is different from the exponent found for one-species avalanches.\\

\begin{figure*}
	\centering
	\subfloat[][]{
		\includegraphics[width=0.45\textwidth]{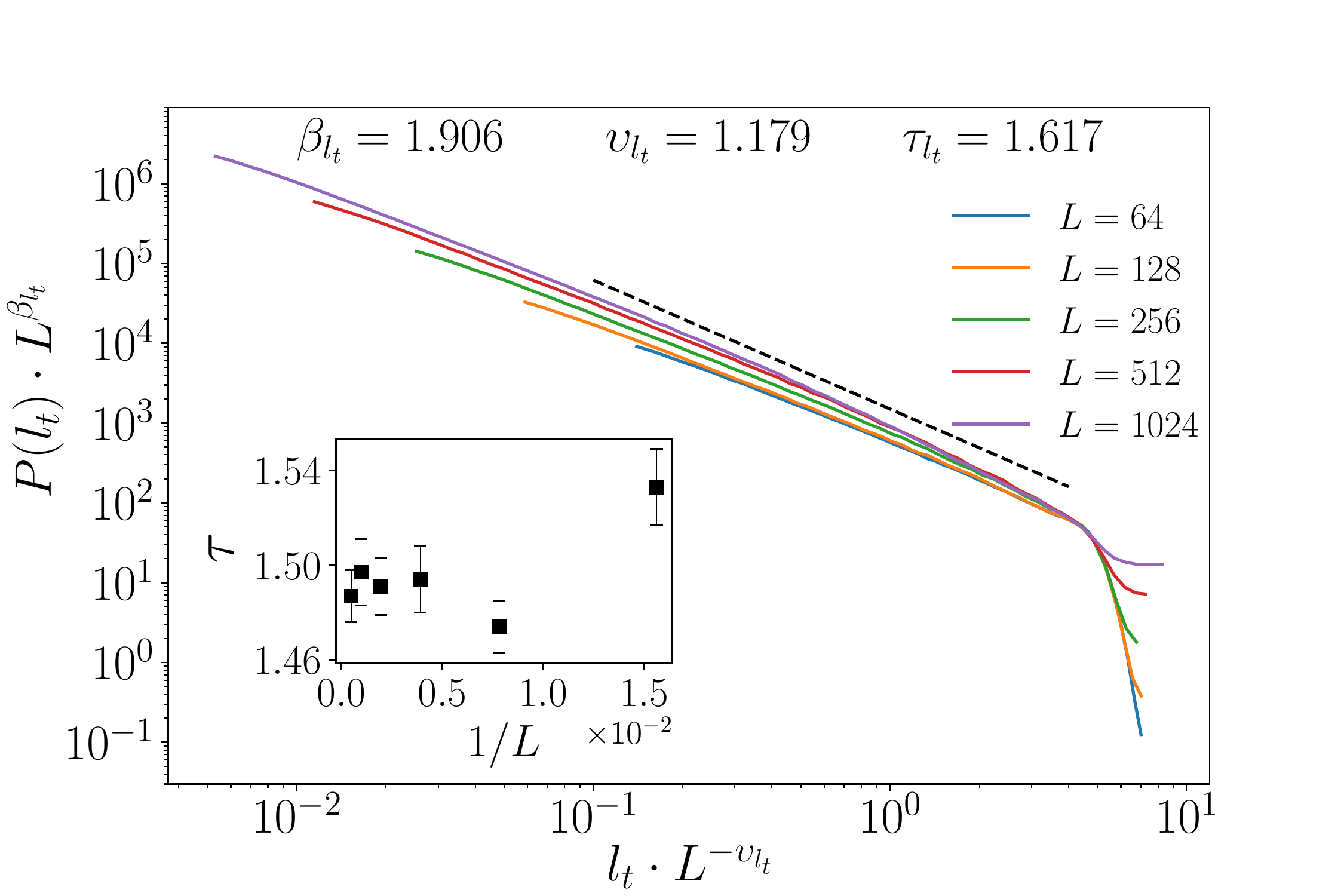}
		\label{fig:P(l_total)}}
	\subfloat[][]{
		\includegraphics[width=0.45\textwidth]{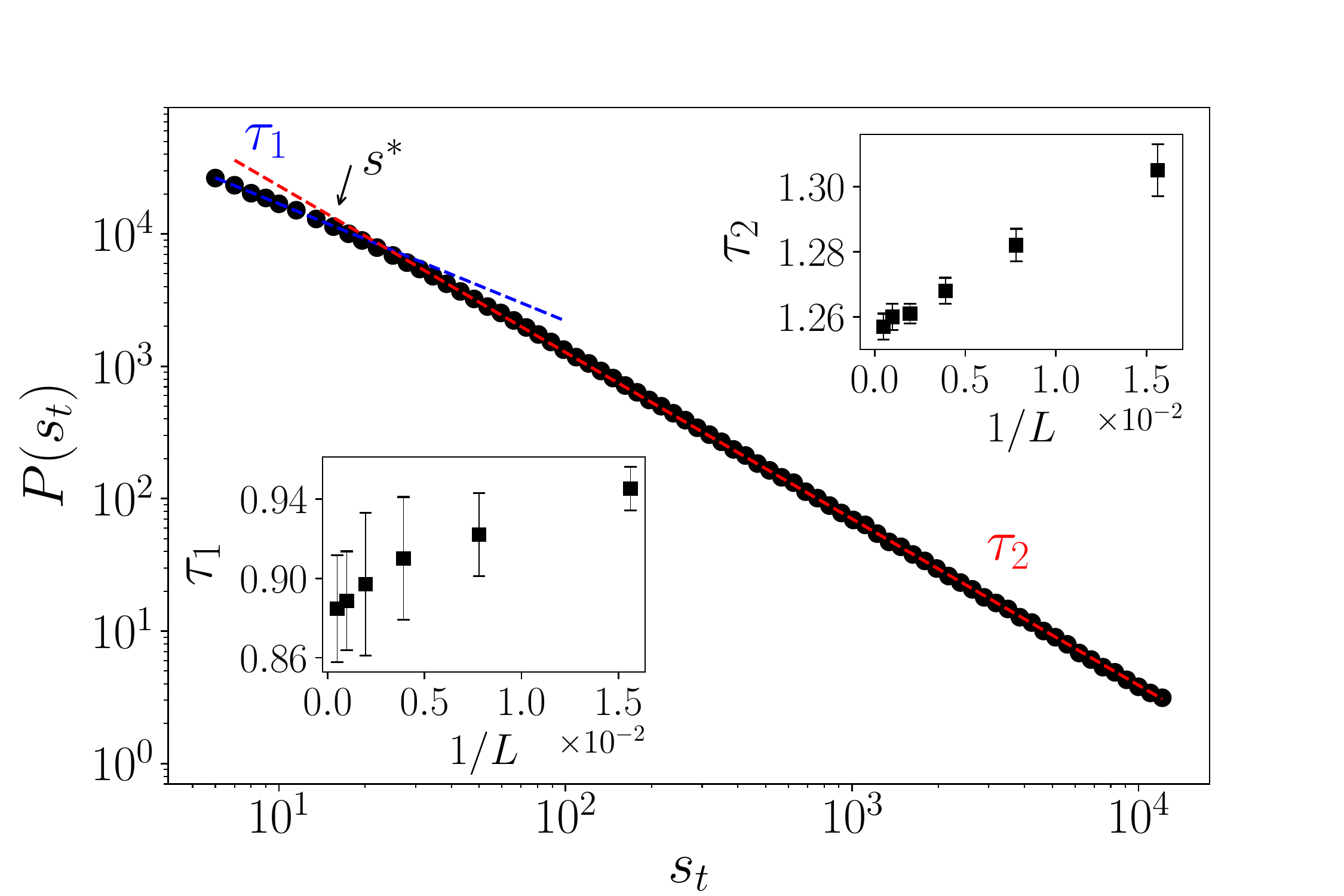}
		\label{fig:P(size_total)}}
	\qquad	
	\subfloat[][]{
		\includegraphics[width=0.45\textwidth]{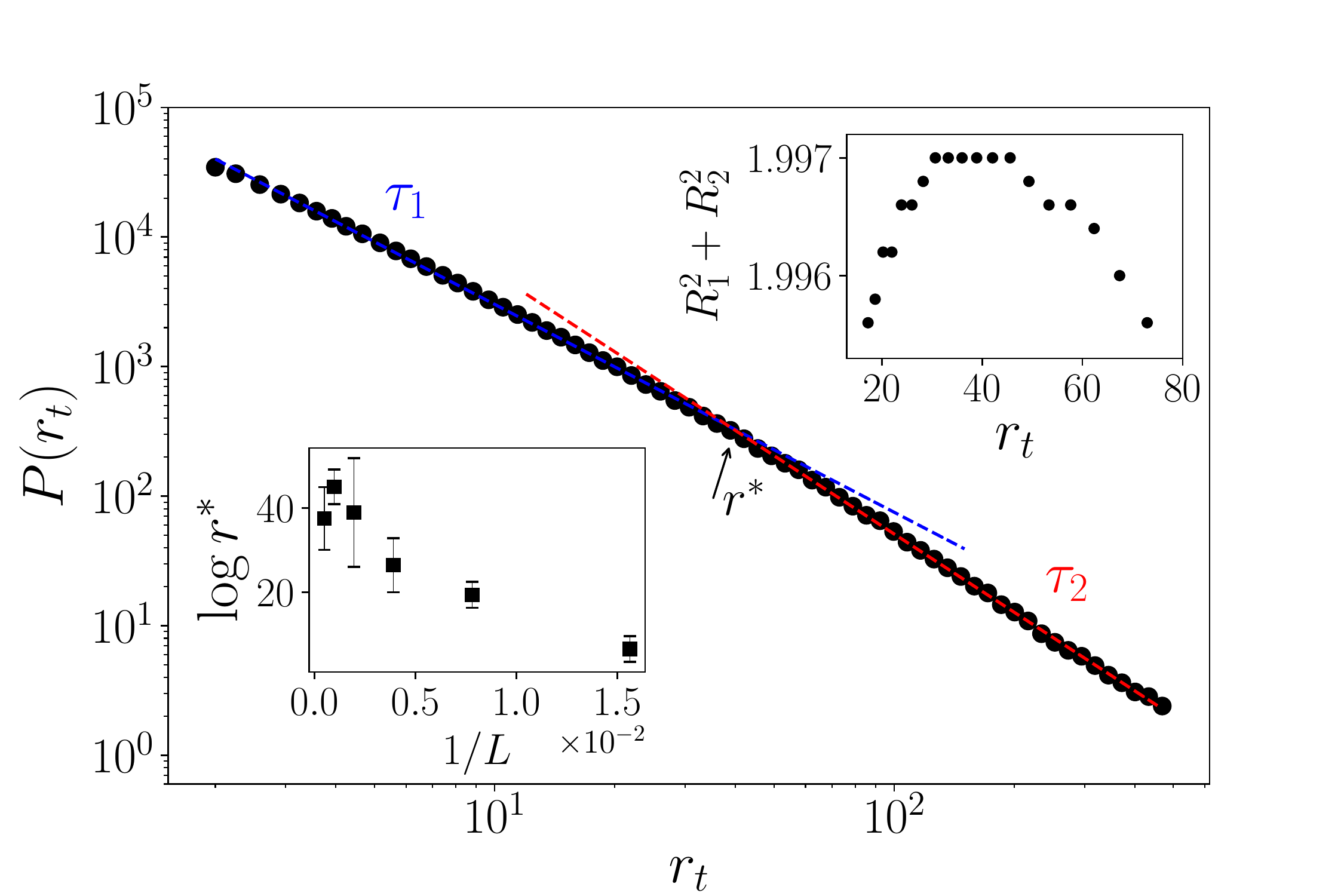}
		\label{fig:r_total2048}}
	\subfloat[][]{
		\includegraphics[width=0.45\textwidth]{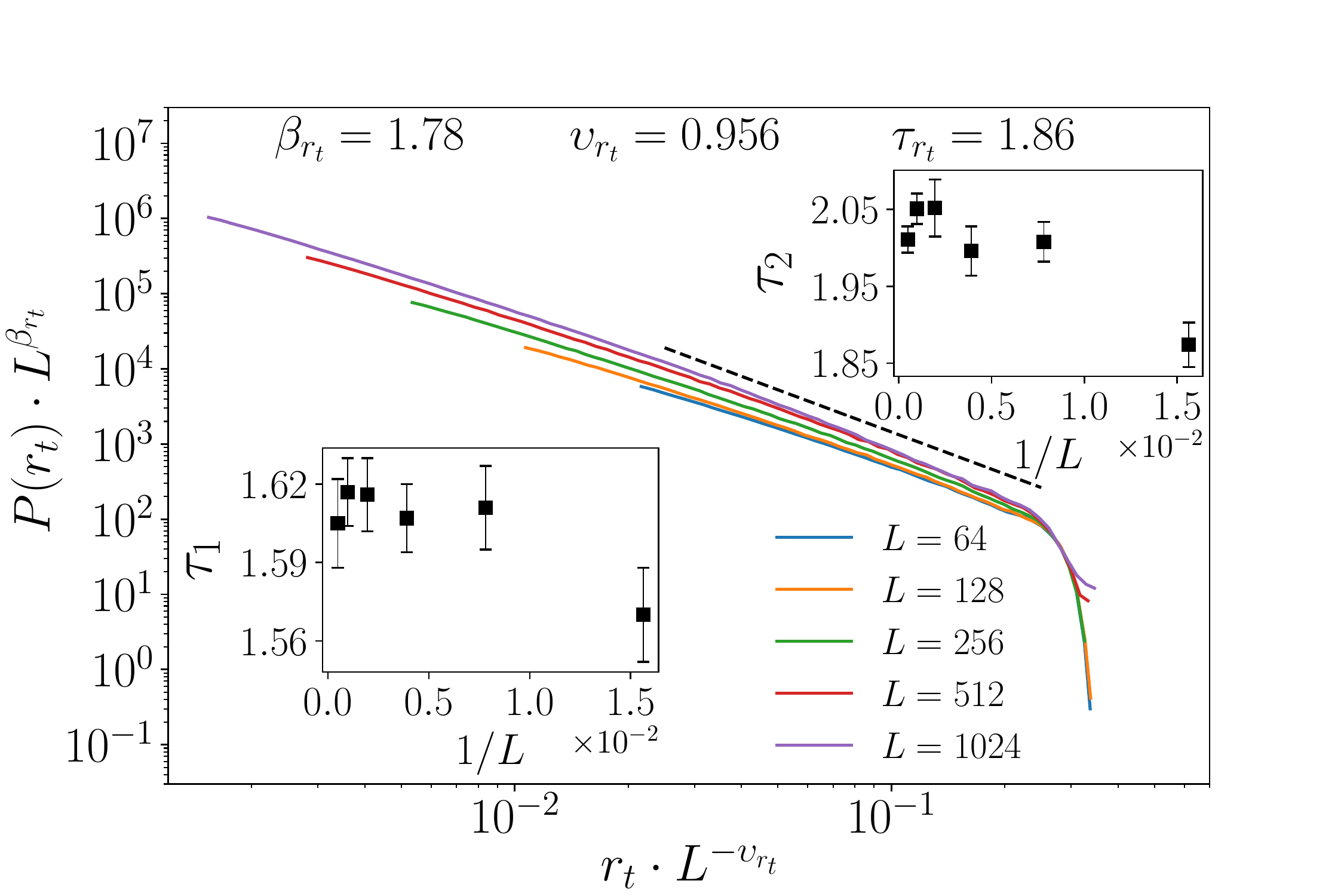}
		\label{fig:P(r_total)}}
	\caption{(Color online): (a) The data collapse for the distribution function of $l_t$ for two-species avalanches, with exponents reported in Table~\ref{tab:exponents2}, and (b) The distribution function of $s_t$, and the crossover point $s^*$, and $\tau_1$ (for small scales, the lower inset) and $\tau_2$ (for large scales, the upper inset) in terms of $1/L$. (c) The distribution function of $r_t$, showing the $L$ dependence of the crossover point $r^*$ (lower inset), and $R_1^2+R_2^2$ (upper inset). (d)  The data collapse for the distribution function of $r_t$, with exponents reported in Table~\ref{tab:exponents2}. }
	\label{Fig:results2}
\end{figure*}

The data collapse for the distribution function for various observables for two-species avalanches are shown in Fig.~\ref{Fig:results2}. The exponents are summarized in TABLE~\ref{tab:exponents2}, which shows deviation from the data that was presented in TABLE~\ref{tab:exponents} for the small scale regime, whereas for large scales regime the results are compatible.   

\begin{table}
	\caption{The exponents $\beta$, $\nu$, $\tau_1$, and $\tau_2$ for $m$, $s$, $l$, and $r$ corresponding to the two-species avalanches.}
	\begin{tabular}{c | c | c | c | c }
		\hline quantity  & $m$ & $s$ & $l$  & $r$ \\
		\hline $\tau_1(L\rightarrow \infty)$  & $0.95\pm 0.05$ & $0.90\pm 0.05$ & $--$  & $1.61\pm 0.05$ \\
		\hline $\tau_2(L\rightarrow \infty)$ & $1.32\pm 0.02$ & $1.25\pm 0.03$ & $1.50\pm 0.03$  & $2.0\pm 0.1$ \\
		\hline $\beta$   & $-$ & $-$ & $1.90\pm 0.05$  & $1.78\pm 0.03$ \\
		\hline $\nu$     & $-$ & $-$ & $1.18\pm 0.03$  & $0.95\pm 0.03$ \\
		\hline
	\end{tabular}
	\label{tab:exponents2}
\end{table}

The spanning avalanche probability (SCP) is defined as the probability that an avalanche percolates, i.e. connects opposite boundaries. In the original BTW-sandpile model, at the mean field level, let $p$ be the probability that a site is minimally stable one, i.e. the site that becomes stable under a single stimulation. $p$ is negligibly small at the beginning of the simulation, and grows as the average height increases. But the average height cannot grow beyond the threshold, the point at which the system relaxes and giant avalanches (the avalanche which touches the boundary) emerge to decrease the average height. At this point the system is self-organized into a critical state. Actually this point is observed when $h_{av}\simeq 3.2$, at which the cluster of minimally stable sites percolates. This process is independent of the number of transported grains in each toppling, i.e. when we let only one grain to pass to the neighbors in one toppling. The probability of forming percolating avalanches is proportional to the probability of giant cluster of minimally stable sites.\\

\begin{figure}
	\centerline{\includegraphics[scale=.38]{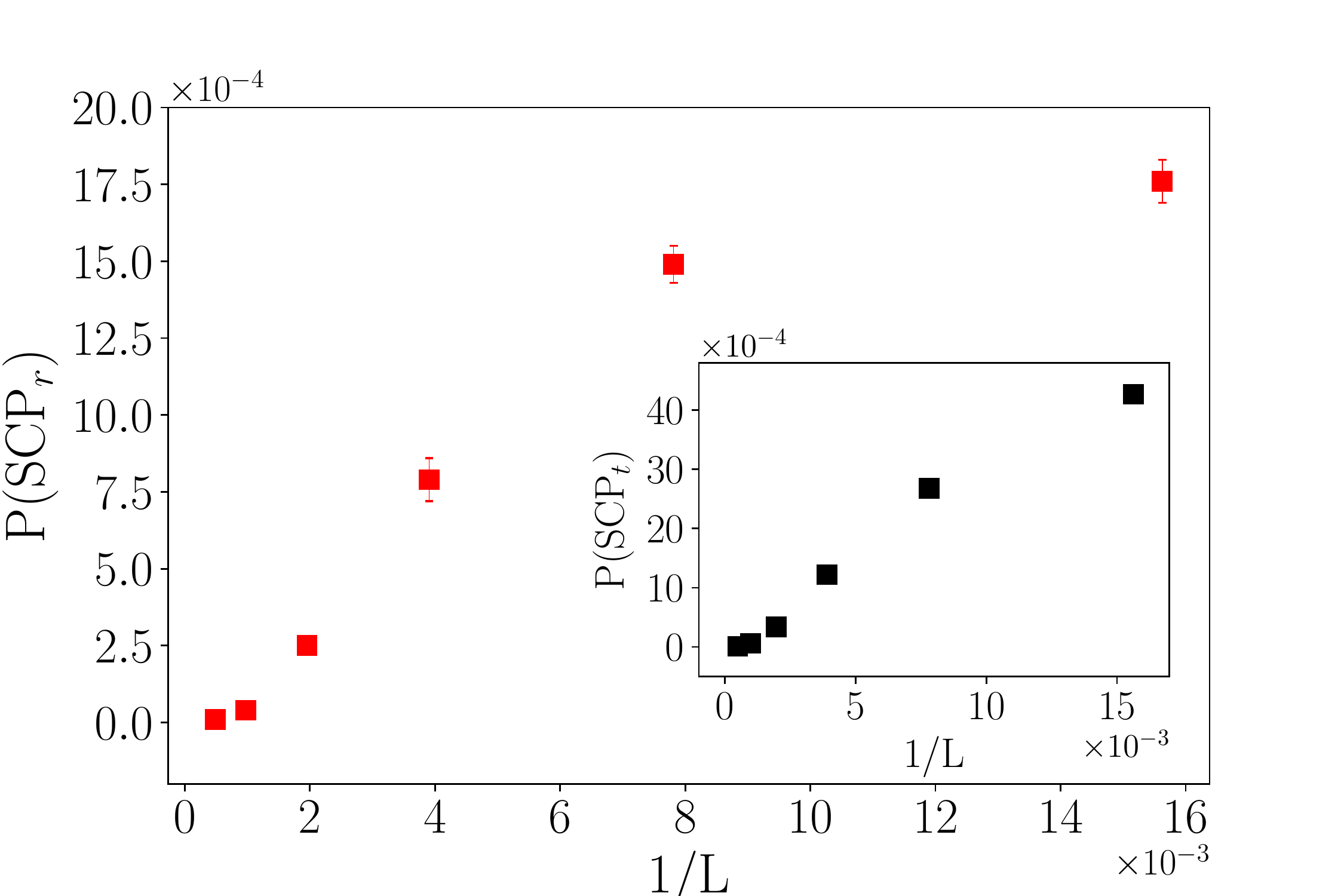}}
	\caption{The spanning cluster probability (SCP) of avalanches as a function of inverse system size $L^{-1}$ for one-species (the main panel) and for two-species (total) avalanches (inset), all extrapolating to zero as $L^{-1}\rightarrow 0$. For small $L^{-1}$ regime, the dependence is nearly linear with slope $0.20\pm 0.05$ for one-species avalanches, and $0.37\pm 0.05$ for two-species avalanches.}
	\label{fig:SCP}
\end{figure}

One expects that this function tends to zero for infinite lattices, however the shape of this dependence is important for comparing it with invasion percolation. The question is: what is the fraction of avalanches that are spanning for lattices of size L? This function is shown in Fig.~\ref{fig:SCP}, which reveals that the spanning cluster probability (SCP) linearly decreases with $\frac{1}{L}$ for large enough systems for both red and total avalanches with different slops. As expected for fixed system size, SCP$_{\text{two-species}}>$ SCP$_{\text{one-species}}$. This is understood in the context of invasion phenomenon: when grains a species trigger avalanches of the other species. In that case, the chance that two-species avalanche percolate is obviously larger than one-species avalanches. 

\section{Concluding Remarks}\label{concl}
We introduced and studied a two-species sandpile model, inspired by multiphase flow in porous media. Different from the BTW model, in the presence of two species, the set of sites that topple in the same relaxation of one of the species is not necessarily connected. Thus, avalanches of one species might trigger avalanches of the other species, what resembles the invasion process observed in porous media. \\

We show that the dynamics is characterized by two different regimes, for small and large avalanches, respectively. While the statistics of the avalanches for the second regime is consistent with what was previously found for the BTW model, the values of the exponents for small avalanches are significantly different, e.g. the fractal dimension of the external perimeter of one species avalanche and the exponent of their size in the small scale regime are $D_f^{(1)}=1.47\pm 0.02$ and $\tau_s^{(1)}=0.95\pm 0.03$. The large scale properties dominate at the thermodynamic limit. We reveal also that the spanning cluster probability (SCP) vanishes in the thermodynamic limit $\frac{1}{L}\rightarrow 0$.\\

\section*{Acknowledgement}
NA acknowledges financial support from the Portuguese Foundation for Science and Technology (FCT) under Contracts nos. PTDC/FIS-MAC/28146/2017 (LISBOA-01-0145-FEDER-028146) , UIDB/00618/2020, and UIDP/00618/2020.

\bibliography{refs}

\begin{thebibliography}{21}%
\makeatletter
\providecommand \@ifxundefined [1]{%
 \@ifx{#1\undefined}
}%
\providecommand \@ifnum [1]{%
 \ifnum #1\expandafter \@firstoftwo
 \else \expandafter \@secondoftwo
 \fi
}%
\providecommand \@ifx [1]{%
 \ifx #1\expandafter \@firstoftwo
 \else \expandafter \@secondoftwo
 \fi
}%
\providecommand \natexlab [1]{#1}%
\providecommand \enquote  [1]{``#1''}%
\providecommand \bibnamefont  [1]{#1}%
\providecommand \bibfnamefont [1]{#1}%
\providecommand \citenamefont [1]{#1}%
\providecommand \href@noop [0]{\@secondoftwo}%
\providecommand \href [0]{\begingroup \@sanitize@url \@href}%
\providecommand \@href[1]{\@@startlink{#1}\@@href}%
\providecommand \@@href[1]{\endgroup#1\@@endlink}%
\providecommand \@sanitize@url [0]{\catcode `\\12\catcode `\$12\catcode
  `\&12\catcode `\#12\catcode `\^12\catcode `\_12\catcode `\%12\relax}%
\providecommand \@@startlink[1]{}%
\providecommand \@@endlink[0]{}%
\providecommand \url  [0]{\begingroup\@sanitize@url \@url }%
\providecommand \@url [1]{\endgroup\@href {#1}{\urlprefix }}%
\providecommand \urlprefix  [0]{URL }%
\providecommand \Eprint [0]{\href }%
\providecommand \doibase [0]{http://dx.doi.org/}%
\providecommand \selectlanguage [0]{\@gobble}%
\providecommand \bibinfo  [0]{\@secondoftwo}%
\providecommand \bibfield  [0]{\@secondoftwo}%
\providecommand \translation [1]{[#1]}%
\providecommand \BibitemOpen [0]{}%
\providecommand \bibitemStop [0]{}%
\providecommand \bibitemNoStop [0]{.\EOS\space}%
\providecommand \EOS [0]{\spacefactor3000\relax}%
\providecommand \BibitemShut  [1]{\csname bibitem#1\endcsname}%
\let\auto@bib@innerbib\@empty
\bibitem [{\citenamefont {Wilkinson}\ and\ \citenamefont
  {Willemsen}(1983)}]{wilkinson1983invasion}%
  \BibitemOpen
  \bibfield  {author} {\bibinfo {author} {\bibfnamefont {D.}~\bibnamefont
  {Wilkinson}}\ and\ \bibinfo {author} {\bibfnamefont {J.~F.}\ \bibnamefont
  {Willemsen}},\ }\href@noop {} {\bibfield  {journal} {\bibinfo  {journal}
  {Journal of Physics A: Mathematical and General}\ }\textbf {\bibinfo {volume}
  {16}},\ \bibinfo {pages} {3365} (\bibinfo {year} {1983})}\BibitemShut
  {NoStop}%
\bibitem [{\citenamefont {Glass}\ and\ \citenamefont
  {Yarrington}(1996)}]{glass1996simulation}%
  \BibitemOpen
  \bibfield  {author} {\bibinfo {author} {\bibfnamefont {R.}~\bibnamefont
  {Glass}}\ and\ \bibinfo {author} {\bibfnamefont {L.}~\bibnamefont
  {Yarrington}},\ }\href@noop {} {\bibfield  {journal} {\bibinfo  {journal}
  {Geoderma}\ }\textbf {\bibinfo {volume} {70}},\ \bibinfo {pages} {231}
  (\bibinfo {year} {1996})}\BibitemShut {NoStop}%
\bibitem [{\citenamefont {Sheppard}\ \emph {et~al.}(1999)\citenamefont
  {Sheppard}, \citenamefont {Knackstedt}, \citenamefont {Pinczewski},\ and\
  \citenamefont {Sahimi}}]{sheppard1999invasion}%
  \BibitemOpen
  \bibfield  {author} {\bibinfo {author} {\bibfnamefont {A.~P.}\ \bibnamefont
  {Sheppard}}, \bibinfo {author} {\bibfnamefont {M.~A.}\ \bibnamefont
  {Knackstedt}}, \bibinfo {author} {\bibfnamefont {W.~V.}\ \bibnamefont
  {Pinczewski}}, \ and\ \bibinfo {author} {\bibfnamefont {M.}~\bibnamefont
  {Sahimi}},\ }\href@noop {} {\bibfield  {journal} {\bibinfo  {journal}
  {Journal of Physics A: Mathematical and General}\ }\textbf {\bibinfo {volume}
  {32}},\ \bibinfo {pages} {L521} (\bibinfo {year} {1999})}\BibitemShut
  {NoStop}%
\bibitem [{\citenamefont {Prat}(1995)}]{prat1995isothermal}%
  \BibitemOpen
  \bibfield  {author} {\bibinfo {author} {\bibfnamefont {M.}~\bibnamefont
  {Prat}},\ }\href@noop {} {\bibfield  {journal} {\bibinfo  {journal}
  {International journal of multiphase flow}\ }\textbf {\bibinfo {volume}
  {21}},\ \bibinfo {pages} {875} (\bibinfo {year} {1995})}\BibitemShut
  {NoStop}%
\bibitem [{\citenamefont {Wettstein}\ \emph {et~al.}(2012)\citenamefont
  {Wettstein}, \citenamefont {Wittel}, \citenamefont {Ara{\'u}jo},
  \citenamefont {Lanyon},\ and\ \citenamefont
  {Herrmann}}]{wettstein2012invasion}%
  \BibitemOpen
  \bibfield  {author} {\bibinfo {author} {\bibfnamefont {S.~J.}\ \bibnamefont
  {Wettstein}}, \bibinfo {author} {\bibfnamefont {F.~K.}\ \bibnamefont
  {Wittel}}, \bibinfo {author} {\bibfnamefont {N.~A.}\ \bibnamefont
  {Ara{\'u}jo}}, \bibinfo {author} {\bibfnamefont {B.}~\bibnamefont {Lanyon}},
  \ and\ \bibinfo {author} {\bibfnamefont {H.~J.}\ \bibnamefont {Herrmann}},\
  }\href@noop {} {\bibfield  {journal} {\bibinfo  {journal} {Physica A:
  Statistical Mechanics and its Applications}\ }\textbf {\bibinfo {volume}
  {391}},\ \bibinfo {pages} {264} (\bibinfo {year} {2012})}\BibitemShut
  {NoStop}%
\bibitem [{\citenamefont {Najafi}\ \emph {et~al.}(2016)\citenamefont {Najafi},
  \citenamefont {Ghaedi},\ and\ \citenamefont
  {Moghimi-Araghi}}]{najafi2016water}%
  \BibitemOpen
  \bibfield  {author} {\bibinfo {author} {\bibfnamefont {M.}~\bibnamefont
  {Najafi}}, \bibinfo {author} {\bibfnamefont {M.}~\bibnamefont {Ghaedi}}, \
  and\ \bibinfo {author} {\bibfnamefont {S.}~\bibnamefont {Moghimi-Araghi}},\
  }\href@noop {} {\bibfield  {journal} {\bibinfo  {journal} {Physica A:
  Statistical Mechanics and its Applications}\ }\textbf {\bibinfo {volume}
  {445}},\ \bibinfo {pages} {102} (\bibinfo {year} {2016})}\BibitemShut
  {NoStop}%
\bibitem [{\citenamefont {Najafi}(2014)}]{najafi2014bak1}%
  \BibitemOpen
  \bibfield  {author} {\bibinfo {author} {\bibfnamefont {M.}~\bibnamefont
  {Najafi}},\ }\href@noop {} {\bibfield  {journal} {\bibinfo  {journal}
  {Physics Letters A}\ }\textbf {\bibinfo {volume} {378}},\ \bibinfo {pages}
  {2008} (\bibinfo {year} {2014})}\BibitemShut {NoStop}%
\bibitem [{\citenamefont {Blunt}(2001)}]{blunt2001flow}%
  \BibitemOpen
  \bibfield  {author} {\bibinfo {author} {\bibfnamefont {M.~J.}\ \bibnamefont
  {Blunt}},\ }\href@noop {} {\bibfield  {journal} {\bibinfo  {journal} {Current
  opinion in colloid \& interface science}\ }\textbf {\bibinfo {volume} {6}},\
  \bibinfo {pages} {197} (\bibinfo {year} {2001})}\BibitemShut {NoStop}%
\bibitem [{\citenamefont {Ara{\'u}jo}(2013)}]{araujo2013getting}%
  \BibitemOpen
  \bibfield  {author} {\bibinfo {author} {\bibfnamefont {N.~A.}\ \bibnamefont
  {Ara{\'u}jo}},\ }\href@noop {} {\bibfield  {journal} {\bibinfo  {journal}
  {Physics}\ }\textbf {\bibinfo {volume} {6}},\ \bibinfo {pages} {90} (\bibinfo
  {year} {2013})}\BibitemShut {NoStop}%
\bibitem [{\citenamefont {Bak}\ \emph {et~al.}(1988)\citenamefont {Bak},
  \citenamefont {Tang},\ and\ \citenamefont {Wiesenfeld}}]{BTW1988Self}%
  \BibitemOpen
  \bibfield  {author} {\bibinfo {author} {\bibfnamefont {P.}~\bibnamefont
  {Bak}}, \bibinfo {author} {\bibfnamefont {C.}~\bibnamefont {Tang}}, \ and\
  \bibinfo {author} {\bibfnamefont {K.}~\bibnamefont {Wiesenfeld}},\ }\href
  {\doibase 10.1103/PhysRevA.38.364} {\bibfield  {journal} {\bibinfo  {journal}
  {Physical Review A}\ }\textbf {\bibinfo {volume} {38}},\ \bibinfo {pages}
  {364} (\bibinfo {year} {1988})}\BibitemShut {NoStop}%
\bibitem [{\citenamefont {Chin}(2002)}]{Chin2002Quant}%
  \BibitemOpen
  \bibfield  {author} {\bibinfo {author} {\bibfnamefont {W.~C.}\ \bibnamefont
  {Chin}},\ }\href@noop {} {\emph {\bibinfo {title} {Quantitative methods in
  reservoir engineering}}}\ (\bibinfo  {publisher} {Gulf Professional
  Publishing},\ \bibinfo {year} {2002})\BibitemShut {NoStop}%
\bibitem [{\citenamefont {Najafi}\ and\ \citenamefont
  {Ghaedi}(2015)}]{najafi2015geometrical}%
  \BibitemOpen
  \bibfield  {author} {\bibinfo {author} {\bibfnamefont {M.}~\bibnamefont
  {Najafi}}\ and\ \bibinfo {author} {\bibfnamefont {M.}~\bibnamefont
  {Ghaedi}},\ }\href@noop {} {\bibfield  {journal} {\bibinfo  {journal}
  {Physica A: Statistical Mechanics and its Applications}\ }\textbf {\bibinfo
  {volume} {427}},\ \bibinfo {pages} {82} (\bibinfo {year} {2015})}\BibitemShut
  {NoStop}%
\bibitem [{\citenamefont {Bak}\ \emph {et~al.}(1987)\citenamefont {Bak},
  \citenamefont {Tang},\ and\ \citenamefont {Wiesenfeld}}]{Bak1987Self}%
  \BibitemOpen
  \bibfield  {author} {\bibinfo {author} {\bibfnamefont {P.}~\bibnamefont
  {Bak}}, \bibinfo {author} {\bibfnamefont {C.}~\bibnamefont {Tang}}, \ and\
  \bibinfo {author} {\bibfnamefont {K.}~\bibnamefont {Wiesenfeld}},\ }\href
  {\doibase 10.1103/PhysRevLett.59.381} {\bibfield  {journal} {\bibinfo
  {journal} {Physical Review Letters}\ }\textbf {\bibinfo {volume} {59}},\
  \bibinfo {pages} {381} (\bibinfo {year} {1987})}\BibitemShut {NoStop}%
\bibitem [{\citenamefont {Dhar}(1999)}]{dhar1999some}%
  \BibitemOpen
  \bibfield  {author} {\bibinfo {author} {\bibfnamefont {D.}~\bibnamefont
  {Dhar}},\ }\href@noop {} {\bibfield  {journal} {\bibinfo  {journal} {Physica
  A: Statistical Mechanics and its Applications}\ }\textbf {\bibinfo {volume}
  {270}},\ \bibinfo {pages} {69} (\bibinfo {year} {1999})}\BibitemShut
  {NoStop}%
\bibitem [{\citenamefont {Najafi}\ \emph
  {et~al.}(2012{\natexlab{a}})\citenamefont {Najafi}, \citenamefont
  {Moghimi-Araghi},\ and\ \citenamefont {Rouhani}}]{najafi2012avalanche}%
  \BibitemOpen
  \bibfield  {author} {\bibinfo {author} {\bibfnamefont {M.}~\bibnamefont
  {Najafi}}, \bibinfo {author} {\bibfnamefont {S.}~\bibnamefont
  {Moghimi-Araghi}}, \ and\ \bibinfo {author} {\bibfnamefont {S.}~\bibnamefont
  {Rouhani}},\ }\href@noop {} {\bibfield  {journal} {\bibinfo  {journal}
  {Physical Review E}\ }\textbf {\bibinfo {volume} {85}},\ \bibinfo {pages}
  {051104} (\bibinfo {year} {2012}{\natexlab{a}})}\BibitemShut {NoStop}%
\bibitem [{\citenamefont {Najafi}\ \emph
  {et~al.}(2012{\natexlab{b}})\citenamefont {Najafi}, \citenamefont
  {Moghimi-Araghi},\ and\ \citenamefont {Rouhani}}]{Najafi2012Observation}%
  \BibitemOpen
  \bibfield  {author} {\bibinfo {author} {\bibfnamefont {M.}~\bibnamefont
  {Najafi}}, \bibinfo {author} {\bibfnamefont {S.}~\bibnamefont
  {Moghimi-Araghi}}, \ and\ \bibinfo {author} {\bibfnamefont {S.}~\bibnamefont
  {Rouhani}},\ }\href {http://stacks.iop.org/1751-8121/45/i=9/a=095001}
  {\bibfield  {journal} {\bibinfo  {journal} {Journal of Physics A:
  Mathematical and Theoretical}\ }\textbf {\bibinfo {volume} {45}},\ \bibinfo
  {pages} {095001} (\bibinfo {year} {2012}{\natexlab{b}})}\BibitemShut
  {NoStop}%
\bibitem [{\citenamefont {Najafi}(2018)}]{najafi2018coupling}%
  \BibitemOpen
  \bibfield  {author} {\bibinfo {author} {\bibfnamefont {M.}~\bibnamefont
  {Najafi}},\ }\href@noop {} {\bibfield  {journal} {\bibinfo  {journal} {arXiv
  preprint arXiv:1801.08978}\ } (\bibinfo {year} {2018})}\BibitemShut {NoStop}%
\bibitem [{\citenamefont {Hoshen}\ and\ \citenamefont
  {Kopelman}(1976)}]{hoshen1976percolation}%
  \BibitemOpen
  \bibfield  {author} {\bibinfo {author} {\bibfnamefont {J.}~\bibnamefont
  {Hoshen}}\ and\ \bibinfo {author} {\bibfnamefont {R.}~\bibnamefont
  {Kopelman}},\ }\href@noop {} {\bibfield  {journal} {\bibinfo  {journal}
  {Physical Review B}\ }\textbf {\bibinfo {volume} {14}},\ \bibinfo {pages}
  {3438} (\bibinfo {year} {1976})}\BibitemShut {NoStop}%
\bibitem [{\citenamefont {L{\"u}beck}\ and\ \citenamefont
  {Usadel}(1997{\natexlab{a}})}]{Lubeck1997BTW}%
  \BibitemOpen
  \bibfield  {author} {\bibinfo {author} {\bibfnamefont {S.}~\bibnamefont
  {L{\"u}beck}}\ and\ \bibinfo {author} {\bibfnamefont {K.~D.}\ \bibnamefont
  {Usadel}},\ }\href {\doibase 10.1103/PhysRevE.56.5138} {\bibfield  {journal}
  {\bibinfo  {journal} {Physical Review E}\ }\textbf {\bibinfo {volume} {56}},\
  \bibinfo {pages} {5138} (\bibinfo {year} {1997}{\natexlab{a}})}\BibitemShut
  {NoStop}%
\bibitem [{\citenamefont {Glantz}\ \emph {et~al.}(1990)\citenamefont {Glantz},
  \citenamefont {Slinker},\ and\ \citenamefont {Neilands}}]{glantz1990primer}%
  \BibitemOpen
  \bibfield  {author} {\bibinfo {author} {\bibfnamefont {S.~A.}\ \bibnamefont
  {Glantz}}, \bibinfo {author} {\bibfnamefont {B.~K.}\ \bibnamefont {Slinker}},
  \ and\ \bibinfo {author} {\bibfnamefont {T.~B.}\ \bibnamefont {Neilands}},\
  }\href@noop {} {\emph {\bibinfo {title} {Primer of applied regression and
  analysis of variance}}},\ Vol.\ \bibinfo {volume} {309}\ (\bibinfo
  {publisher} {McGraw-Hill New York},\ \bibinfo {year} {1990})\BibitemShut
  {NoStop}%
\bibitem [{\citenamefont {L{\"u}beck}\ and\ \citenamefont
  {Usadel}(1997{\natexlab{b}})}]{Lubeck1997Numerical}%
  \BibitemOpen
  \bibfield  {author} {\bibinfo {author} {\bibfnamefont {S.}~\bibnamefont
  {L{\"u}beck}}\ and\ \bibinfo {author} {\bibfnamefont {K.~D.}\ \bibnamefont
  {Usadel}},\ }\href {\doibase 10.1103/PhysRevE.55.4095} {\bibfield  {journal}
  {\bibinfo  {journal} {Physical Review E}\ }\textbf {\bibinfo {volume} {55}},\
  \bibinfo {pages} {4095} (\bibinfo {year} {1997}{\natexlab{b}})}\BibitemShut
  {NoStop}%
\end{thebibliography}%

\end{document}